%% file: R15MH.tex
\documentclass[noinfoline,stslayout]{imsart}
\usepackage{wrapfig}
\usepackage{natbib}
\usepackage{graphicx}
\usepackage{amsfonts,amssymb,amsmath,amscd,amsthm,latexsym}
\usepackage[]{units}
\usepackage{qtree}
\usepackage[mathletters]{ucs}
\usepackage[utf8x]{inputenc}
\usepackage{algorithm,algorithmic,nicefrac}

\usepackage[usenames]{color}
\usepackage{framed}
\definecolor{shadecolor1}{rgb}{0.80,0.73,0.85}
\definecolor{shadecolor2}{rgb}{0.30,0.73,0.60}

\newenvironment{shaded1}{%
  \MakeFramed {\FrameRestore}}%
 {\endMakeFramed}

\newenvironment{shaded2}{%
  \MakeFramed {\FrameRestore}}%
 {\endMakeFramed}

\newtheorem{alga}{Algorithm}

\newenvironment{algo}
  {\medskip\begin{shaded2}\begin{alga}\begin{sffamily}}
  {\end{sffamily}\end{alga}\end{shaded2}\medskip}

\newtheorem{example}{Example}{\bfseries}{\rmfamily}

\newenvironment{ex}
    {\begin{example}\begin{em}}
    {\end{em}\hfill$\blacktriangleleft$\end{example}}

\DeclareGraphicsExtensions{.pdf,.png,.jpg}

\begin{document}


\begin{frontmatter}
\title{The Metropolis--Hastings algorithm}
\author{C.P. Robert${}^{1,2,3}$}
\affiliation{${}^1$Université Paris-Dauphine, ${}^2$University of Warwick, and ${}^3$CREST}

\begin{abstract}
This article is a self-contained introduction to the Metropolis-Hastings algorithm, this ubiquitous tool for producing
dependent simulations from an arbitrary distribution. The document illustrates the principles of the methodology on simple
examples with R codes and provides entries to the recent extensions of the method.
\end{abstract}

\begin{keyword}
\kwd{Bayesian inference}
\kwd{Markov chains}
\kwd{MCMC methods}
\kwd{Metropolis--Hastings algorithm}
\kwd{intractable density}
\kwd{Gibbs sampler}
\kwd{Langevin diffusion}
\kwd{Hamiltonian Monte Carlo}
\end{keyword}

\end{frontmatter}

\section{Introduction}\label{sec:intro}

There are many reasons why computing an integral like 
$$\mathfrak{I}(h)=\int_\mathcal{X} h(x)\text{d}\pi(x),$$ 
where $\text{d}\pi$ is a probability measure, may prove
intractable, from the shape of the domain $\mathcal{X}$ to the dimension of $\mathcal{X}$ (and $x$), to the complexity
of one of the functions $h$ or $\pi$. Standard numerical methods may be hindered by the same reasons. Similar
difficulties (may) occur when attempting to find the extrema of $\pi$ over the domain $\mathcal{X}$. This is why the
recourse to Monte Carlo methods may prove unavoidable: exploiting the probabilistic nature of $\pi$ and its weighting of
the domain $\mathcal{X}$ is often the most natural and most efficient way to produce approximations to integrals
connected with $\pi$ and to determine the regions of the domain $\mathcal{X}$ that are more heavily weighted by $\pi$.
The Monte Carlo approach \citep{hammersley:handscomb:1964,rubinstein:1981} emerged with computers, at the end of WWII,
as it relies on the ability of producing a large number of realisations of a random variable distributed according to a
given distribution, taking advantage of the stabilisation of the empirical average predicted by the Law of Large
Numbers.  However, producing simulations from a specific distribution may prove near impossible or quite costly and
therefore the (standard) Monte Carlo may also face intractable situations.

An indirect approach to the simulation  of complex distributions and in particular to the curse of dimensionality met by
regular Monte Carlo methods is to use a Markov chain associated with this target
distribution, using Markov chain theory to validate the convergence of the chain to the distribution of interest and the
stabilisation of empirical averages \citep{meyn:tweedie:1994}. It is thus little surprise that
Markov chain Monte Carlo (MCMC) methods have been used for almost as long as the original Monte Carlo techniques, even
though their impact on Statistics has not been truly felt until the very early 1990s. A comprehensive entry about the
history of MCMC methods can be found in \cite{robert:casella:2010}. 

The paper\footnote{I am most grateful to Alexander Ly, Department of Psychological Methods, University of
Amsterdam, for pointing out mistakes in the R code of an earlier version of this paper.} 
is organised as follows: in Section \ref{sec:algo}, we define and justify the Metropolis--Hastings algorithm,
along historical notes about its origin. In Section \ref{sec:Zimp}, we provide details on the implementation and
calibration of the algorithm. A mixture example is processed in Section \ref{sec:ex}. Section \ref{sec:ext} includes
recent extensions of the standard Metropolis--Hastings algorithm, while Section \ref{sec:con} concludes about further
directions for Markov chain Monte Carlo methods when faced with complex models and huge datasets.

\section{The algorithm}\label{sec:algo}

\subsection{Motivations}
\begin{wrapfigure}{R}{0.5\textwidth}
\vspace{-10pt}
\includegraphics[width=0.45\textwidth]{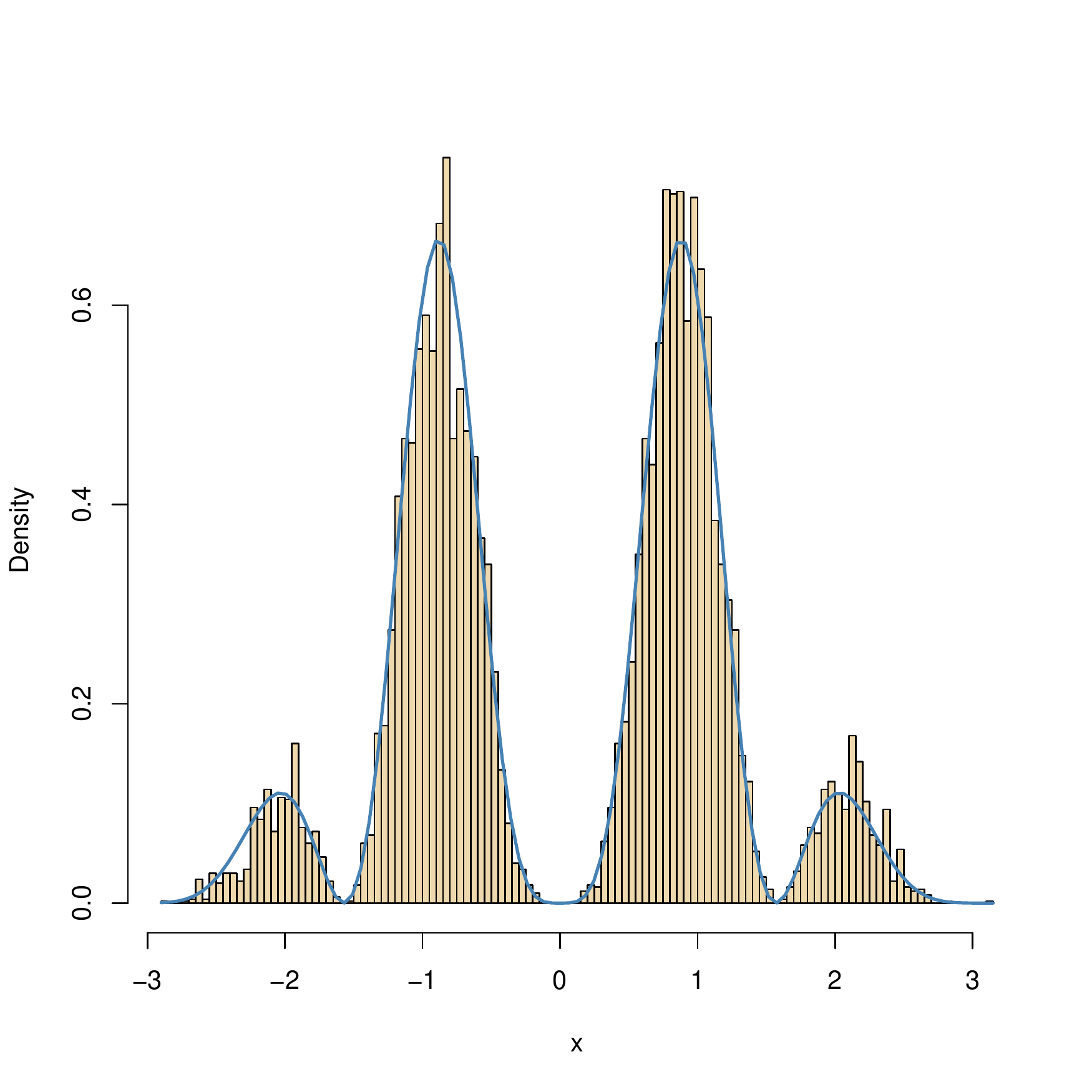}
\vspace{-10pt}
\caption{\label{fig:toysin}
Fit of the histogram of a Metropolis--Hastings sample to its target, for $T=10^4$ iterations, a scale $\alpha=1$, and a
starting value $x^{(1)}=3.14$.}
\vspace{-10pt}
\end{wrapfigure}
Given a probability density $\pi$ called the {\em target}, defined on a state space $\mathcal{X}$, and computable up to
a multiplying constant, $\pi(x)\propto\tilde{\pi}(x)$, the Metropolis--Hastings algorithm, named after
\cite{Metropolis:Rosenbluth:Rosenbluth:Teller:Teller:1953} and \cite{Hastings:1970}, proposes a generic way to 
construct a Markov chain on $\mathcal{X}$ that is ergodic and stationary with respect to $\pi$---meaning that, if
$X^{(t)}\sim\pi(x)$, then $X^{(t+1)}\sim\pi(x)$---and that therefore converges in distribution to $\pi$. While there are
other generic ways of delivering Markov chains associated with an arbitrary stationary distribution, see, e.g.,
\cite{Barker:1965}, the Metropolis--Hastings algorithm is the workhorse of MCMC methods, both for its simplicity and its
versatility, and hence the first solution to consider in intractable situations. The main motivation for using Markov
chains is that they provide shortcuts in cases where generic sampling requires too much effort from the experimenter.
Rather than aiming at the ``big picture" immediately, as an accept-reject algorithm would do
\citep{robert:casella:2009}, Markov chains construct a progressive picture of the target distribution, proceeding by
local exploration of the state space $\mathcal{X}$ until all the regions of interest have been uncovered. An analogy for
the method is the case of a visitor to a museum forced by a general blackout to watch a painting with a small torch. Due
to the narrow beam of the torch, the person cannot get a global view of the painting but can proceed along this painting
until all parts have been seen.\footnote{Obviously, this is only an analogy in that a painting is more than
the sum of its parts!}

\begin{wrapfigure}{R}{0.5\textwidth}
\vspace{-10pt}
\includegraphics[width=0.45\textwidth]{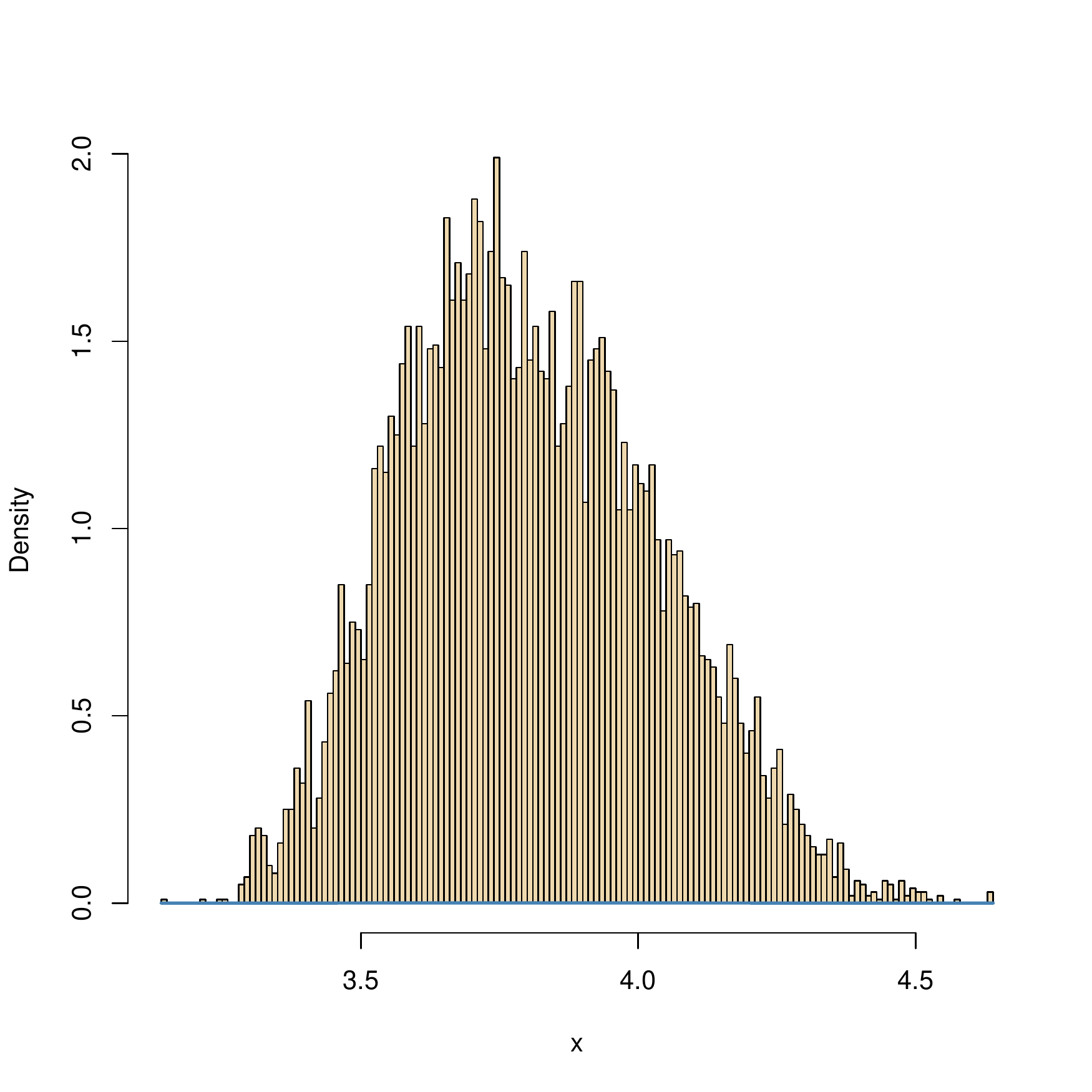}
\vspace{-10pt}
\caption{\label{fig:kapitalsin}
Fit of the histogram of a Metropolis--Hastings sample to its target, for $T=10^4$ iterations, a scale $\alpha=0.1$, and a
starting value $x^{(1)}=3.14$.}
\vspace{-10pt}
\end{wrapfigure}
Before describing the algorithm itself, let us stress the probabilistic foundations of Markov chain Monte Carlo (MCMC)
algorithms: the Markov chain returned by the method, $X^{(1)},X^{(2)},\ldots,X^{(t)},\ldots$ is such that $X^{(t)}$ is
converging to $\pi$.  This means that the chain can be considered as a sample, albeit a dependent sample, and
approximately distributed from $\pi$. Due to the Markovian nature of the simulation, the first values are highly
dependent on the starting value $X^{(1)}$ and usually removed from the sample as {\em burn-in} or {\em warm-up}. While
there are very few settings where {\em the} time when the chain {\em reaches} stationarity can be determined, see, e.g.,
\cite{hobert:robert:2002}, there is no need to look for such an instant since the empirical average
\begin{equation}\label{eq:empimean}
		\hat{\mathfrak{I}}_T(h)=\dfrac{1}{T}\sum_{t=1}^T h(X^{(t)})
\end{equation}
converges almost surely to $\mathfrak{I}(h)$, no matter what the starting value, if the Markov chain is ergodic, i.e.,
forgets about its starting value. This implies that, in theory, simulating a Markov chain is intrinsically equivalent to
a standard i.i.d.~simulation from the target, the difference being in a loss of efficiency, i.e., in the necessity to
simulate more terms to achieve a given variance for the above Monte Carlo estimator. The foundational principle for MCMC
algorithms is thus straightforward, even though the practical implementation of the method may prove delicate or in
cases impossible.

Without proceeding much further into Markov chain theory, we stress that the existence of a stationary distribution for
a chain implies this chain automatically enjoys a strong stability called {\em irreducibility}. Namely, the chain can
move all over the state space, i.e., can eventually reach any region of the state space, no matter its initial value.

\subsection{The algorithm}
The Metropolis--Hastings algorithm associated with a target density $\pi$ requires the choice of a
conditional density $q$ also called {\em proposal or candidate kernel}. The transition from the value of
the Markov chain $(X^{(t)})$ at time $t$ and its value at time $t+1$ proceeds via the following transition step:

\begin{algo}\label{al:mh}
{\bf Metropolis--Hastings}\\
\noindent {\sf Given $X^{(t)}=x^{(t)}$,
\begin{enumerate}
\item Generate $\; Y_t \sim q(y|x^{(t)})$.

\item Take
$$
X^{(t+1)} = \begin{cases}
Y_t & \hbox{{\tt with probability} }\ \rho(x^{(t)},Y_t), \cr
x^{(t)} & \hbox{{\tt with probability} }\ 1 - \rho(x^{(t)},Y_t), \cr
\end{cases}
$$
\noindent {\tt where}
$$
\rho(x,y) = \min \left\{ \dfrac{\tilde{\pi}(y)}{\tilde{\pi}(x)} \;
\dfrac{q(x|y)}{q(y|x)} \;, 1 \right\} \,.
$$
\end{enumerate}
}

\end{algo}

\begin{wrapfigure}{R}{0.5\textwidth}
\vspace{-10pt}
\includegraphics[width=.45\textwidth]{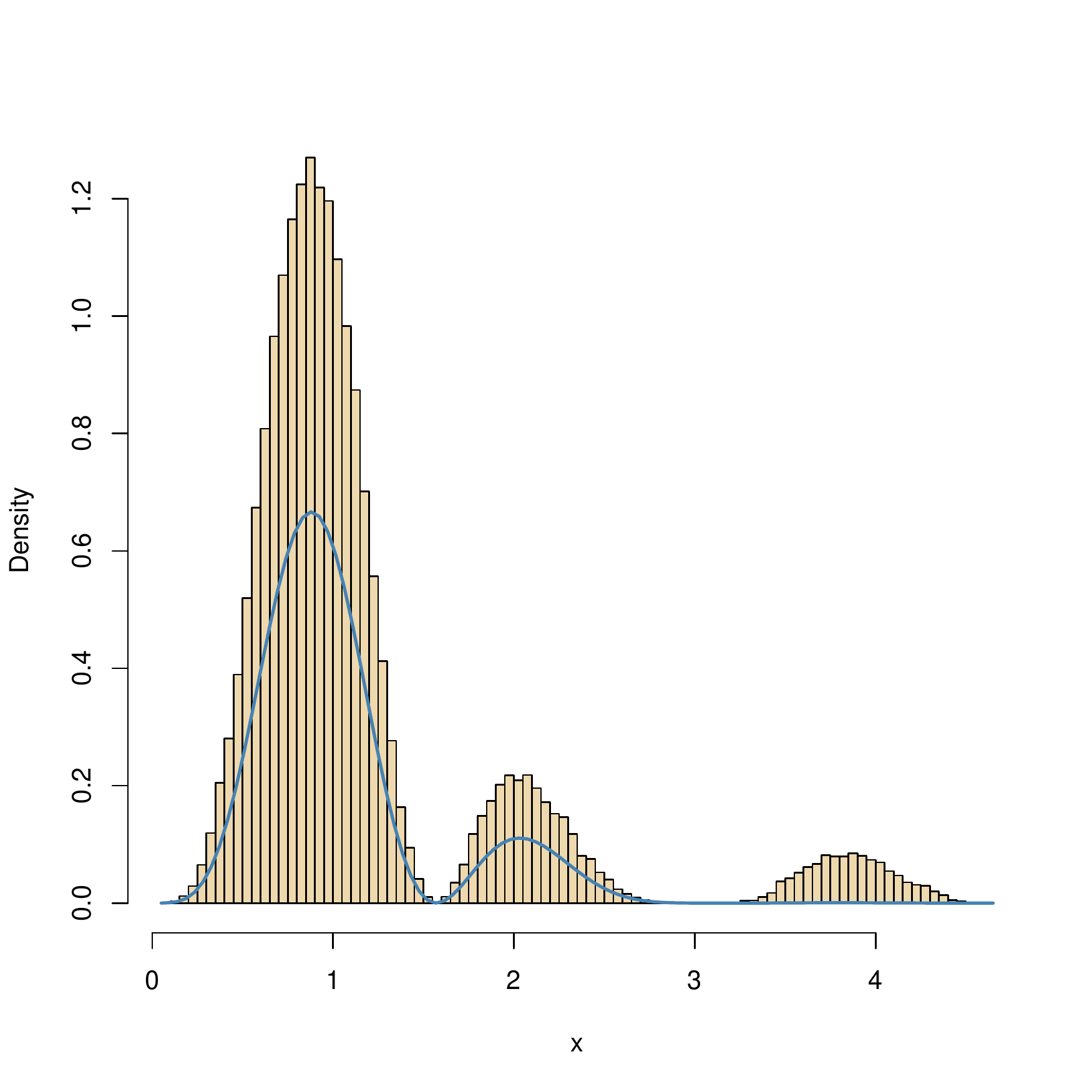}
\vspace{-10pt}
\caption{\label{fig:weaksin}
Fit of the histogram of a Metropolis--Hastings sample to its target, for $T=10^5$ iterations, a scale $\alpha=0.2$, and a
starting value $x^{(1)}=3.14$.}
\vspace{-10pt}
\end{wrapfigure}
Then, as shown in \cite{Metropolis:Rosenbluth:Rosenbluth:Teller:Teller:1953}, this transition preserves the stationary
density $\pi$ if the chain is irreducible, that is, if $q$ has a wide enough support to eventually reach any region of
the state space $\mathcal{X}$ with positive mass under $\pi$. A sufficient condition is that $q$ is positive everywhere.
The very nature of accept-reject step introduced by those authors is therefore sufficient to turn a simulation from an
almost arbitrary proposal density $q$ into a generation that preserves $\pi$ as the stationary distribution. This sounds
both amazing and too good to be true! But it is true, in the theoretical sense drafted above. In practice, the
performances of the algorithm are obviously highly dependent on the choice of the transition $q$, since some choices see
the chain unable to converge in a manageable time.

\subsection{An experiment with the algorithm}

To capture the mechanism behind the algorithm, let us consider an elementary example:

\begin{ex}\label{ex:toysin}
Our target density is a perturbed version of the normal $\mathcal{N}(0,1)$ density, $\varphi(\cdot)$,
$$\tilde\pi(x) = \sin^2(x) \times \sin^2(2x) \times \varphi(x)\,.$$
And our proposal is a uniform $\mathcal{U}(x-\alpha,x+\alpha)$ kernel, 
$$q(y|x) = \frac{1}{2\alpha}\mathbb{I}_{(x-\alpha,x+\alpha)}(y)\,.$$
Implementing this algorithm is straightforward: two functions to define are the target and the transition
\begin{shaded1}\begin{verbatim}
target=function(x){
  sin(x)^2*sin(2*x)^2*dnorm(x)}

metropolis=function(x,alpha=1){
  y=runif(1,x-alpha,x+alpha)
  if (runif(1)>target(y)/target(x)) y=x
  return(y)}
\end{verbatim}\end{shaded1}
\noindent and all we need is a starting value
\begin{shaded1}\begin{verbatim}
T=10^4
x=rep(3.14,T)                                                                           
for (t in 2:T) x[t]=metropolis(x[t-1])                                                   
\end{verbatim}\end{shaded1}
\noindent which results in the histogram of Figure \ref{fig:toysin}, where the target density is properly normalised by a
numerical integration. If we look at the sequence $(x^{(t)})$ returned by the algorithm, it changes values around $5000$
times. This means that one proposal out of two is rejected. If we now change the scale of the uniform to $\alpha=0.1$,
the chain $(x^{(t)})$ takes more than $9000$ different values, however the histogram in Figure \ref{fig:kapitalsin}
shows a poor fit to the target in that only one mode is properly explored. The proposal lacks the power to move the
chain far enough to reach the other parts of the support of $\pi(\cdot)$. A similar behaviour occurs when we start at
$0$. A last illustration of the possible drawbacks of using this algorithm is shown on Figure \ref{fig:weaksin}: when
using the scale $\alpha=0.2$ the chain is slow in exploring the support, hence does not reproduce the correct shape of
$\pi$ after $T=10^5$ iterations.
\end{ex}

From this example, we learned that some choices of proposal kernels work well to recover the shape of the target
density, while others are poorer, and may even fail altogether to converge. Details about the implementation of the
algorithm and the calibration of the proposal $q$ are detailed in Section \ref{sec:Zimp}.

\subsection{Historical interlude}\label{sec:stor}

The initial geographical localisation of the MCMC algorithms is the nuclear research laboratory in Los Alamos, New
Mexico, which work on the hydrogen bomb eventually led to the derivation Metropolis algorithm in the early 1950s. 
What can be reasonably seen as the first MCMC algorithm is indeed the Metropolis algorithm,
published by \cite{Metropolis:Rosenbluth:Rosenbluth:Teller:Teller:1953}. Those algorithms are thus contemporary with
the standard Monte Carlo method, developped by Ulam and von Neumann in the late 1940s. (Nicolas Metropolis is also
credited with suggesting the name ``Monte Carlo``, see \citealp{Eckhardt:1987}, and published the very first Monte Carlo
paper, see \citealp{metropolis:ulam:1949}.) This Metropolis algorithm, while used in physics, was only generalized by
\cite{Hastings:1970} and \cite{Peskun:1973,Peskun:1981} towards statistical applications, as a method apt to overcome
the curse of dimensionality penalising regular Monte Carlo methods. Even those later generalisations and the work of
Hammersley, Clifford, and Besag in the 1970's did not truly impact the statistical community until \cite{Geman:Geman:1984}
experimented with the Gibbs sampler for image processing, \cite{Tanner:Wong:1987} created a form of Gibbs sampler for
latent variable models and \cite{Gelfand:Smith:1990} extracted the quintessential aspects of Gibbs sampler to turn it
into a universal principle and rekindle the appeal of the Metropolis--Hastings algorithm for Bayesian computation and
beyond.

\section{Implementation details}\label{sec:Zimp}

\begin{wrapfigure}{R}{0.5\textwidth}
\vspace{-10pt}
\includegraphics[width=.45\textwidth]{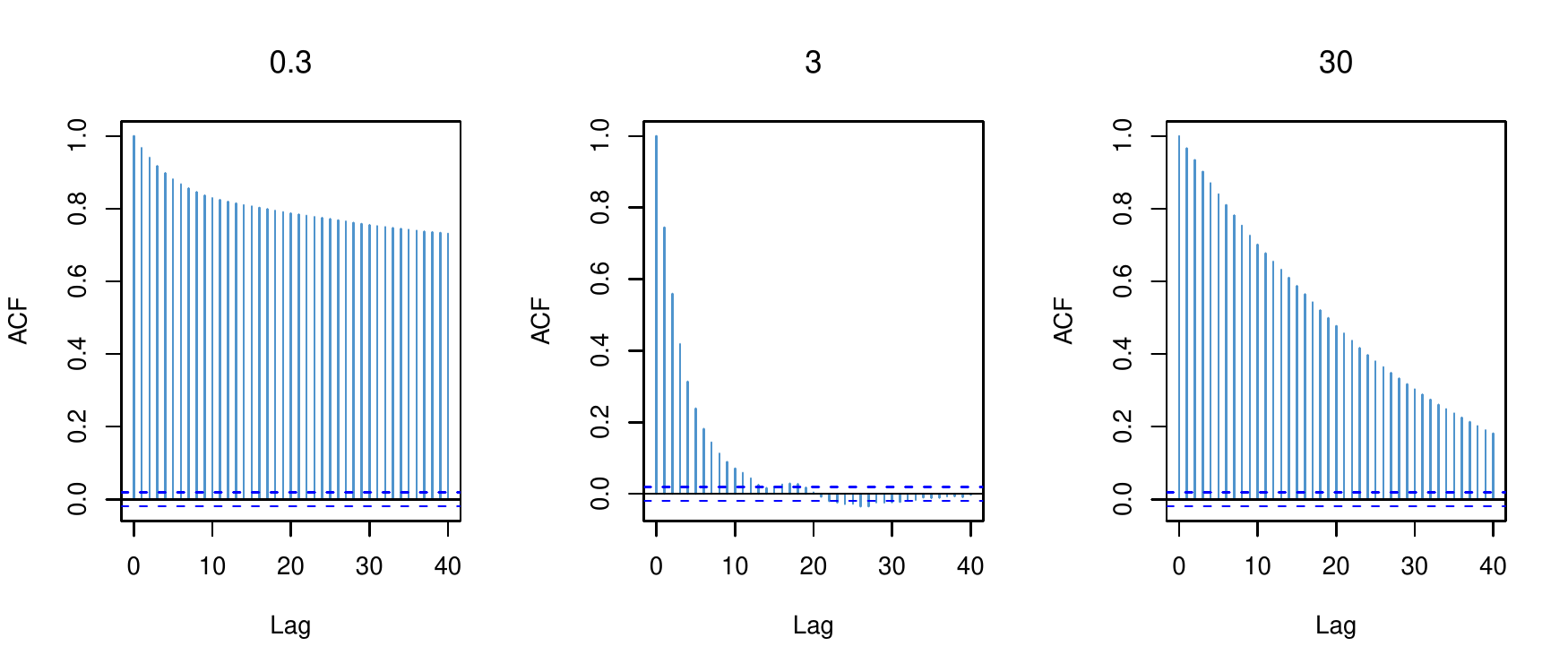}
\caption{\label{fig:acfsin}
Comparison of the acf of three Markov chains corresponding to scales $\alpha=0.3,3,30$, for $T=10^4$ iterations, and a
starting value $x^{(1)}=3.14$.}
\vspace{-10pt}
\end{wrapfigure}

When working with a Metropolis--Hastings algorithm, the generic nature of Algorithm \ref{al:mh} is as much an hindrance
as a blessing in that the principle remains valid for almost every choice of the proposal $q$. It thus does not give
indications about the calibration of this proposal. For instance, in Example \ref{ex:toysin}, the method is valid for
all choices of $\alpha$ but the comparison of the histogram of the outcome with the true density shows that $\alpha$ has
a practical impact on the convergence of the algorithm and hence on the number of iterations it requires. Figure
\ref{fig:acfsin} illustrates this divergence in performances via the autocorrelation graphs of three chains produced by
the R code in Example \ref{ex:toysin} for three highly different values of $\alpha=0.3,3,30$. It shows why $\alpha=3$
should be prefered to the other two values in that each value of the Markov chain contains ``more"  information in that
case. The fundamental difficulty when using the Metropolis--Hastings algorithm is in uncovering which calibration is
appropriate without engaging into much experimentation or, in other words, in an as automated manner as possible.

A (the?) generic version of the Metropolis--Hastings algorithm is the {\em random walk Metropolis--Hastings algorithm},
which exploits as little as possible knowledge about the target distribution, proceeding instead in a local if often
myopic manner. To achieve this, the proposal distribution $q$ aims at a {\em local} exploration of the neighborhood of
the current value of the Markov chain, i.e., simulating the proposed value $Y_t$ as
\begin{displaymath}
Y_t=X^{(t)} + \varepsilon_t,
\end{displaymath}
where $\varepsilon_t$ is a random perturbation with distribution $g$, for instance a uniform distribution as in Example
\ref{ex:toysin} or a normal distribution. If we call {\em random walk Metropolis--Hastings algorithms} all the cases
when $g$ is symmetric, the acceptance probability in Algorithm \ref{al:mh} gets simplified into
$$
\rho(x,y) = \min \left\{ \dfrac{\tilde{\pi}(y)}{\tilde{\pi}(x)}\;, 1 \right\}\,.
$$
 
While this probability is independent of the scale of the proposal $g$, we just saw that the performances of the
algorithm are quite dependent on such quantities. In order to achieve an higher degree of efficiency, i.e., towards a
decrease of the Monte Carlo variance, \cite{Roberts:Gelman:Gilks:1997} studied a formal Gaussian setting aiming at the
ideal acceptance rate. Indeed, Example \ref{ex:toysin} showed that acceptance rates that are either ``too high" or ``too
low" slow down the convergence of the Markov chain. They then showed that the ideal variance in the proposal is twice
the variance of the target or, equivalently, that the acceptance rate should be close to $\nicefrac{1}{4}$.
While this rule is only an indication (in the sense that it was primarily designed for a specific and asymptotic
Gaussian environment), it provides a golden rule for the default calibration of random walk Metropolis--Hastings
algorithms.

\subsection{Effective sample size}

We now consider the alternative of the {\em effective sample size} for comparing and calibrating
MCMC algorithms. Even for a stationary Markov chain, using $T$ iterations does not amount to simulating an iid sample
from $\pi$ that would lead to the same variability. Indeed, the empirical average \eqref{eq:empimean} cannot be
associated with the standard variance estimator
$$
\hat\sigma^2_T = \frac{1}{T-1} \sum_{t=1}^T \left( h(X^{(t)}) - \hat{\mathfrak{I}}_T(h)_T \right)^2
$$
due to the correlations amongst the $X^{(t)}$'s. In this setting, the {\em
effective sample size} is defined as the correction factor $\tau_T$ such that
$\hat\sigma^2_T/\tau_T$ is the variance of the empirical average \eqref{eq:empimean}).
This quantity can be computed as in \cite{Geweke:1992} and \cite{Heidelberger:Welch:1983} by
$$
  \tau_T = T / \kappa(h)\,,
$$
where $\kappa(h)$ is the autocorrelation associated with the sequence $h(X^{(t)})$,
$$
  \kappa(h) = 1 + 2\,\sum_{t=1}^\infty
  \hbox{corr}\left( h(X^{(0)}),h(X^{(t)}) \right)\,,
$$
estimated by \verb+spectrum0+ and \verb+effectiveSize+ from the R library \verb+coda+, via the spectral density at zero.
A rough alternative is to rely on subsampling, as in \cite{robert:casella:2009}, so that $X^{(t+G)}$ is approximately
independent from $X^{(t)}$. The lag $G$ is possibly determined in R via the autocorrelation function \verb=autocorr=.

\begin{ex} {\bf (Example \ref{ex:toysin} continued)} We can compare the Markov chains obtained with $\alpha=0.3,3,30$
against those two criteria:
\begin{shaded1}\begin{verbatim}
> autocor(mcmc(x))
            [,1] 	[,2]	[,3]
Lag 0  1.0000000 1.0000000 1.0000000
Lag 1  0.9672805 0.9661440 0.9661440
Lag 5  0.8809364 0.2383277 0.8396924
Lag 10 0.8292220 0.0707092 0.7010028
Lag 50 0.7037832 -0.033926 0.1223127
> effectiveSize(x)
      [,1]       [,2]       [,3]
  33.45704 1465.66551  172.17784 
\end{verbatim} \end{shaded1}
This shows how much comparative improvement is brought by the value $\alpha=3$, but also that even this quasi-optimal
case is far from an i.i.d.
setting.
\end{ex}

\subsection{In practice}

In practice, the above tools of ideal acceptance rate and of higher effective sample size give goals for calibrating
Metropolis--Hastings algorithms. This means comparing a range of values of the parameters involved in the proposal and
selecting the value that achieves the highest target for the adopted goal. For a multidimensional parameter, global
maximisation run afoul of the curse of dimensionality as exploring a grid of possible values quickly becomes impossible.
The solution to this difficulty stands in running partial optimisations, with simulated (hence controlled) data, 
for instance setting all parameters but one fixed to the values used for the simulated data. If this is not possible, 
optimisation of the proposal parameters can be embedded in a Metropolis-within-Gibbs\footnote{The
Metropolis-within-Gibbs algorithm aims at simulating a multidimensional distribution by successively simulating from
some of the associated conditional distributions---this is the Gibbs part---and by using one Metropolis--Hastings step
instead of an exact simulation scheme from this conditional, with the same validation as the original Gibbs sampler.}
since for each step several values of the corresponding parameter can be compared via the Metropolis--Hastings
acceptance probability.

\begin{wrapfigure}{R}{.5\textwidth}
\vspace{-10pt}
\includegraphics[width=.45\textwidth]{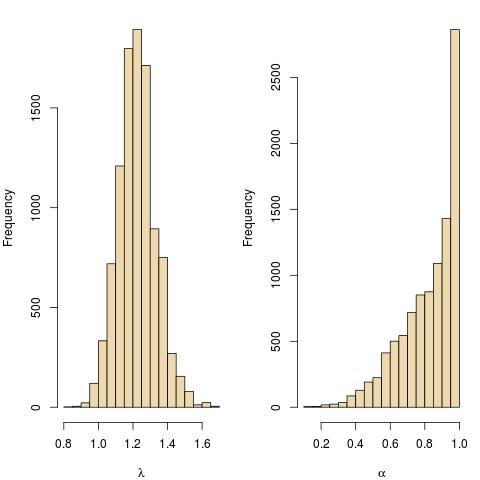}
\vspace{-10pt}
\caption{\label{fig:2dMet}
Output of a two-dimensional random walk Metropolis--Hastings algorithm for 123 observations from a Poisson distribution
with mean $1$, under the assumed model of a mixture between Poisson and Geometric distributions.}
\vspace{-10pt}
\end{wrapfigure}
We refer the reader to Chapter 8 of \cite{robert:casella:2009} for more detailed descriptions of the calibration of MCMC
algorithms, including the use of adaptive mecchanisms. Indeed, calibration is normaly operated in a warm-up stage since,
otherwise, if one continuously tune an MCMC algorithm according to its past outcome, the algorithm stops being Markovian.
In order to preserve convergence in an adaptive MCMC algorithm, the solution found in the literature for this difficulty
is to progressively tone/tune down the adaptive aspect. For instance, \cite{roberts:rosenthal:2006} propose
a {\em diminishing adaptation} condition that states that the distance between two consecutive Markov kernels must
uniformly decrease to zero. For instance, a random walk proposal that relies on the empirical variance of the
past sample as suggested in \cite{haario:sacksman:tamminen:1999} does satisfy this condition. An alternative proposed by
\cite{roberts:rosenthal:2006} proceeds by tuning the scale of a random walk for each component against the
acceptance rate, which is the solution implemented in the \verb+amcmc+ package developed by \cite{rosenthal:2007}.

\section{Illustration}\label{sec:ex}

\begin{wrapfigure}{R}{.5\textwidth}
\vspace{-10pt}
\includegraphics[width=.45\textwidth]{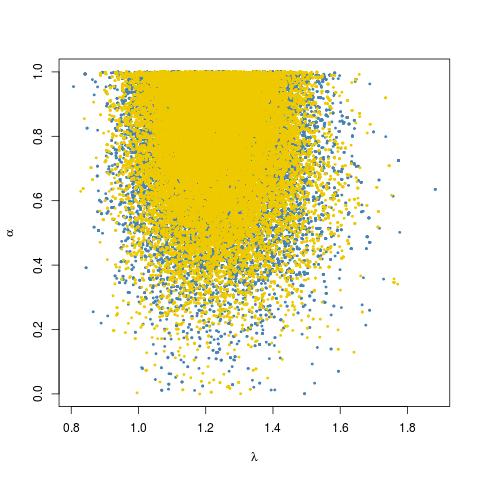}
\vspace{-10pt}
\caption{\label{fig:12Met}
Output of a Metropolis-within-Gibbs {\em (blue)} and of a two-dimensional {\em (gold)} random walk Metropolis--Hastings
algorithm for 123 observations from a Poisson distribution with mean $1$, under the assumed model of a mixture between
Poisson and Geometric distributions.}
\end{wrapfigure}

\cite{kamary:mengersen:x:rousseau:2014} consider the special case of a mixture of a Poisson and of a Geometric
distributions with the same mean parameter $\lambda$:
$$
\alpha \mathcal{P}(\lambda) +(1-\alpha) \mathcal{G}eo(\nicefrac{1}{1+\lambda})\,,
$$
where $\lambda>0$ and $0\le\alpha\le 1$. Given $n$ observations $(x_1,\ldots,x_n)$ and a prior decomposed into 
$\pi(\lambda)\propto\nicefrac{1}{\lambda}$ and $\pi(\alpha)\propto[\alpha(1-\alpha)]^{a_0-1}$, $a_0=0.5$ being the
default value, the likelihood function is available in closed form as
$$
\prod_{i=1}^n \left\{\alpha \frac{\lambda^{x_i}}{x_i!}\exp\{-\lambda\} +(1-\alpha)\lambda^{x_i}(1+\lambda)^{-x_i-1} \right\}
$$
where $s_n=x_1+\ldots+x_n$.  In R code, this translates as
\begin{shaded1}\begin{verbatim}
likelihood=function(x,lam,alp){
  prod(alp*dpois(x,lam)+(1-alp)*dgeom(x,lam/(1+lam)))}
posterior=function(x,lam,alp){
  sum(log(alp*dpois(x,lam)+(1-alp)*dgeom(x,1/(1+lam))))-
      log(lam)+dbeta(alp,.5,.5,log=TRUE)}
\end{verbatim}\end{shaded1}
If we want to build a Metropolis--Hastings algorithm that simulates from the associated posterior, the proposal can
proceed by either proposing a joint move on $(\alpha,\lambda)$ or moving one parameter at a time in a
Metropolis-within-Gibbs fashion. In the first case, we can imagine a random walk in two dimensions,
$$
\alpha'\sim\mathcal{E}(\epsilon\alpha,\epsilon(1-\alpha))\,,\quad
\lambda'\sim \mathcal{LN}(\log(\lambda),\delta(1+\log(\lambda)^2))\,,\qquad \epsilon\,,\delta>0
$$
with an acceptance probability
$$
\dfrac{\pi(\alpha',\lambda'|x)q(\alpha,\lambda|\alpha',\lambda')}
{\pi(\alpha,\lambda|x)q(\alpha,\lambda|\alpha',\lambda')}\wedge 1\,.
$$
The Metropolis--Hastings R code would then be
\begin{shaded1}\begin{verbatim}
metropolis=function(x,lam,alp,eps=1,del=1){
  prop=c(exp(rnorm(1,log(lam),sqrt(del*(1+log(lam)^2)))), 
         rbeta(1,1+eps*alp,1+eps*(1-alp)))
  rat=posterior(x,prop[1],prop[2])-posterior(x,lam,alp)+
         dbeta(alp,1+eps*prop[2],1+eps*(1-prop[2]),log=TRUE)-
         dbeta(prop[2],1+eps*alp,1+eps*(1-alp),log=TRUE)+
         dnorm(log(lam),log(prop[1]),
               sqrt(del*(1+log(prop[1])^2)),log=TRUE)-
         dnorm(log(prop[1]),log(lam),
               sqrt(del*(1+log(lam)^2)),log=TRUE)+
         log(prop[1]/lam)
  if (log(runif(1))>rat) prop=c(lam,alp)
  return(prop)}
\end{verbatim}\end{shaded1}
\noindent where the ratio \verb+prop[1]/lam+ in the acceptance probability is just
the Jacobian for the log-normal transform. Running the following R code
\begin{shaded1}\begin{verbatim}
T=1e4
x=rpois(123,lambda=1)
para=matrix(c(mean(x),runif(1)),nrow=2,ncol=T)
like=rep(0,T)
for (t in 2:T){
   para[,t]=metropolis(x,para[1,t-1],para[2,t-1],eps=.1,del=.1)
   like[t]=posterior(x,para[1,t],para[2,t])}
\end{verbatim}\end{shaded1}
\noindent then produced the histograms of Figure \ref{fig:2dMet}, after toying with the values of $\epsilon$ and
$\delta$ to achieve a large enough average acceptance probability, which is provided by
\verb+length(unique(para[1,]))/T+
The second version of the Metropolis--Hastings algorithm we can test is to separately modify $\lambda$ by a random walk
proposal, test whether or not it is acceptable, and repeat with $\alpha$: the R code is then very similar to the above
one:
\begin{shaded1}\begin{verbatim}
metropolis=function(x,lam,alp,eps=1,del=1){
  prop=exp(rnorm(1,log(lam),sqrt(del*(1+log(lam)^2))))
  rat=posterior(x,prop,alp)-posterior(x,lam,alp)+
         dnorm(log(lam),log(prop[1]),
               sqrt(del*(1+log(prop[1])^2)),log=TRUE)-
         dnorm(log(prop[1]),log(lam),
               sqrt(del*(1+log(lam)^2)),log=TRUE)+
         log(prop/lam)
  if (log(runif(1))>rat) prop=lam
  qrop=rbeta(1,1+eps*alp,1+eps*(1-alp))
  rat=posterior(x,prop,qrop)-posterior(x,prop,alp)+
         dbeta(alp,1+eps*qrop,1+eps*(1-qrop),log=TRUE)-
         dbeta(qrop,1+eps*alp,1+eps*(1-alp),log=TRUE)
  if (log(runif(1))>rat) qrop=alp
  return(c(prop,qrop))}
\end{verbatim}\end{shaded1}
\begin{wrapfigure}{R}{.5\textwidth}
\vspace{-10pt}
\includegraphics[width=.45\textwidth]{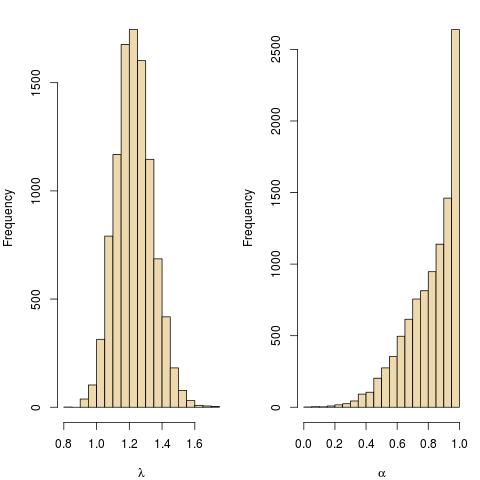}
\vspace{-10pt}
\caption{\label{fig:1dMet}
Output of a Metropolis-within-Gibbs random walk Metropolis--Hastings algorithm for 123 observations from a Poisson
distribution with mean $1$, under the assumed model of a mixture between Poisson and Geometric distributions.}
\vspace{-10pt}
\end{wrapfigure}
In this special case, both algorithms thus return mostly equivalent outcomes, with a slightly more dispersed output in
the case of the Metropolis-within-Gibbs version (Figure \ref{fig:12Met}). In a more general perspective, calibrating
random walks in multiple dimensions may prove unwieldly, especially with large dimensions, while the
Metropolis-within-Gibbs remains manageable. One drawback of the later is common to all Gibbs implementations, namely
that it induces higher correlations between the components, which means a slow convergence of the chain (and in extreme
cases no convergence at all).

\section{Extensions}\label{sec:ext}

\subsection{Langevin algorithms}
An extension of the random walk Metropolis--Hastings algorithm is based on the Langevin diffusion solving
$$
\text{d} X_t = \nicefrac{1}{2} \nabla \log \pi(X_t) \text{d} t + \text{d}B_t\,,
$$
where $B_t$ is the standard Brownian motion and $\nabla f$ denotes the gradient of $f$, 
since this diffusion has $ \pi $ as its stationary and limiting distribution. 
The algorithm is based on a discretised version of the above, namely
$$
Y_{n+1} | X_n \sim \mathcal{N}\left( x + \nicefrac{h}{2} \nabla \log \pi (x), h^{\nicefrac{1}{2}}\,I_d\right)\,,
$$
for a discretisation step $h$, which is used as a proposed value for $X_{n+1}$, and accepted with
the standard Metropolis--Hastings probability \citep{roberts:tweedie:1995}. This new proposal took the name of
Metropolis adjusted Langevin algorithms (hence MALA).
While computing (twice) the gradient of $\pi$ at each iteration requires extra time, there is strong support for doing
so, as MALA algorithms do provide noticeable speed-ups in convergence for most problems. Note that $\pi(\cdot)$ only
needs to be known up to a multiplicative constant because of the log transform. 

\subsection{Particle MCMC}
Another extension of the Metropolis--Hastings algorithm is the particle MCMC (or {\em pMCMC}), developed by
\cite{andrieu:doucet:holenstein:2010}. While we cannot provide an introduction to particle filters here, see, e.g.,
\cite{delmoral:doucet:jasra:2006}, we want to point out the appeal of this approach in state space models like hidden
Markov models (HMM). This innovation is similar to the pseudo-marginal algorithm approach of
\cite{beaumont:2003,andrieu:roberts:2009}, taking advantage of the auxiliary variables exploited by
particle filters. 

In the case of an HMM, i.e., where a {\em latent} Markov chain $x_{0:T}$ with density
$$
p_0(x_0|\theta)p_1(x_1|x_0,\theta)\cdots p_T(x_T|x_{T-1},\theta)\,,
$$
is associated with an {\em observed} sequence $y_{1+T}$ such that
$$
y_{1+T}|X_{1:T},\theta \sim \prod_{i=1}^T q_i(y_i|x_i,\theta)\,,
$$
pMCMC applies as follows. At every iteration $t$, a value $\theta^\prime$ of the parameter
$\theta\sim\mathfrak{h}(\theta|\theta^{(t)})$ is proposed, followed by a new value of the latent series
$x_{0:T}^\prime$ generated from a particle filter approximation of $p(x_{0:T}|\theta^\prime,y_{1:T})$.
As the particle filter produces in addition \citep{delmoral:doucet:jasra:2006} an unbiased estimator 
of the marginal posterior of $y_{1:T}$, $\hat{q}(y_{1:T}|\theta^\prime)$, this estimator can be directly
included in the Metropolis--Hastings ratio
$$
\dfrac{\hat{q}(y_{1:T}|\theta^\prime)\pi(\theta^\prime)\mathfrak{h}(\theta^{(t)}|\theta^\prime)}{\hat{q}(y_{1:T}|\theta)\pi(\theta^{(t)})\mathfrak{h}(\theta^\prime|\theta^{(t)})}
\wedge 1\,.
$$
The validation of this substitution follows from the general argument of \cite{andrieu:roberts:2009} for pseudo-marginal
techniques, even though additional arguments are required to establish that all random variables used therein are
accounted for (see \citealp{andrieu:doucet:holenstein:2010} and \citealp{wilkinson:2011}). We however stress that the
general validation of those algorithm as converging to the joint posterior does not proceed from pseudo-marginal arguments.
An extension of pMCMC called $\text{SMC}^2$ that approximates the sequential filtering distribution is proposed in \cite{chopin:jacob:papaspiliopoulos:2013}.

\subsection{Pseudo-marginals}\label{sub:psudos}

As illustrated by the previous section, there are many settings where computing the target density $\pi(\cdot)$ is
impossible. Another example is made of doubly intractable likelihoods \citep{murray:etal:2006}, when the likelihood
function contains a term that is intractable, for instance $\ell(\theta|x)\propto g(x|\theta)$ with an
intractable normalising constant
$$
\mathfrak{Z}(\theta) = \int_\mathcal{X}  g(x|\theta) \,\text{d}x\,.
$$
This phenomenon is quite common in graphical models, as for instance for the Ising model
\citep{murray:nested:potts,moller:etal:2006}. Solutions based on auxiliary variables have 
been proposed \citep[see, e.g.,][]{murray:etal:2006,moller:etal:2006}, but they may prove difficult to calibrate.

In such settings, \cite{andrieu:roberts:2009} developped an approach based on an idea of \cite{beaumont:2003},
designing a valid Metropolis--Hastings algorithm that substitutes the intractable target $\pi(\cdot|x)$ with 
an unbiased estimator. A slight change to the Metropolis--Hastings acceptance ratio ensures that the
stationary density of the corresponding Markov chain is still equal to the target $\pi$. Indeed, provided
$\hat\pi(\theta|z)$ is an unbiased estimator of $\pi(\theta)$ when $z\sim q(\cdot|\theta)$, it is rather
straightforward to check that the acceptance ratio
$$
\dfrac{\hat\pi(\theta^*|z^*)}{\hat\pi(\theta|z)}\,\dfrac{q(\theta^*,\theta)
q(z|\theta)}{q(\theta,\theta^*)q(z^*|\theta^*)}
$$
preserves stationarity with respect to an extended target (see \cite{andrieu:roberts:2009}) for details)  when 
$z^*\sim q(\cdot|\theta)$, and $\theta^*|\theta\sim q(\theta,\theta^*)$. \cite{andrieu:vihola:2012} propose an
alternative validation via auxiliary weights used in the unbiased estimation, assuming the unbiased estimator (or the
weight) is generated conditional on the proposed value in the original Markov chain. 
The performances of pseudo-marginal solutions depend on the quality of the estimators $\hat{\pi}$ and are always poorer
than when using the exact target $\pi$. In particular, improvements can be found by using multiple
samples of $z$ to estimate $\pi$ \citep{andrieu:vihola:2012}.

\section{Conclusion and new directions}\label{sec:con}

The Metropolis--Hastings algorithm is to be understood as a default or off-the-shelf solution, meaning that (a) it rarely
achieves optimal rates of convergence \citep{mengersen:tweedie:1996} and may get into convergence difficulties if
improperly calibrated but (b) it can be combined with other solutions as a baseline solution, offering further local or
more rarely global exploration to a taylored algorithm. Provided {\em reversibility} is preserved, it is indeed valid to
mix several MCMC algorithms together, for instance picking one of the kernels at random or following a cycle
\citep{tierney:1994,robert:casella:2004}. Unless a proposal is relatively expensive to compute or to implement, it
rarely hurts to add an extra kernel into the MCMC machinery.  

This is not to state that the Metropolis--Hastings algorithm is the ultimate solution to all simulation and stochastic
evaluation problems. For one thing, there exist settings where the intractability of the target is such that no
practical MCMC solution is available. For another thing, there exist non reversible versions of those algorithms, like
Hamiltonian (or hybrid) Monte Carlo (HMC) \citep{duane:etal:1987,neal:2013,betancourt:byrne:linvingstone:girolami:2014}.
This method starts by creating a completly artificial variable $p$, inspired by the momentum in physics, and a joint
distribution on $(q,p)$ which energy---minus the log-density---is defined by the Hamiltonian
$$
H(q,p) = -\log\pi(q)+p^\text{T} M^{-1} p / 2\,,
$$
where $M$ is the so-called mass matrix. The second part in the target is called the kinetic energy, still by analogy
with mechanics. When the joint vector $(q,p)$ is driven by Hamilton's equations
\begin{align*}
\dfrac{\text{d}q}{\text{d}t} &= \dfrac{\partial H}{\partial p} =  \dfrac{\partial H}{\partial p} = M^{-1} p\\
\dfrac{\text{d}p}{\text{d}t} &= -\dfrac{\partial H}{\partial q} =  \dfrac{\partial \log\pi}{\partial q}\\
\end{align*} 
this dynamics preserves the joint distribution with density $\exp -H(p,q)$. If we could simulate {\em exactly} 
from this joint distribution of $(q,p)$, a sample from $\pi(q)$ would be a by-product.
In practice, the equation is only solved approximately and hence requires a Metropolis--Hastings correction.
Its practical implementation is called the {\em leapfrog approximation} \citep{neal:2013,girolami:2011} as it relies 
on a small discretisation step $\epsilon$, updating $p$ and $q$ via a modified Euler's method called the leapfrog 
that is reversible and preserves volume as well. This discretised update can be repeated for an arbitrary number of steps.

The appeal of HMC against other MCMC solutions is that the value of the Hamiltonian changes very little during the 
Metropolis step, while possibly producing a very different value of $q$. Intuitively, moving along level sets in the
augmented space is almost energy-free, but if those move proceeds far enough, the Markov chain on $q$ can reach distant
regions, thus avoid the typical local nature of regular MCMC algorithms. This strengthe explains in part why a
statistical software like STAN \citep{stan-software:2014} is mostly based on HMC moves.

As a last direction for new MCMC solutions, let us point out the requirements set by Big Data, i.e., in settings where
the likelihood function cannot be cheaply evaluated for the entire dataset. See, e.g.,
\cite{scott:etal:2013,wang:dunson:2013}, for recent entries on different parallel ways of handling massive datasets, and
\cite{brockwell:2006,strid:2010,banterle:etal:2015} for delayed and prefetching MCMC techniques that avoid considering
the entire likelihood at once.

\input R15MH.bbl

\end{document}

%% file: R15MH.bbl
\hyphenation{Post-Script Sprin-ger}